# Wealth inequality and utility: Effect evaluation of redistribution and consumption morals using the macro-econophysical coupled approach

**Takeshi Kato*[1], Yosuke Tanabe[2], and Mohammad Rezoanul Hoque[3]**

*[1]Hitachi Kyoto University Laboratory, Kyoto University, Kyoto, Japan*
*[2]Next Research, Research and Development Group, Hitachi, Ltd., Tokyo, Japan*
*[3]Department of Economics, Texas Tech University, Lubbock, TX, USA*

## Summary

Reducing wealth inequality and increasing utility are critical issues. This study reveals the effects of redistribution and consumption morals on wealth inequality and utility. To this end, we present a novel approach that couples the dynamic model of capital, consumption, and utility in macroeconomics with the interaction model of joint business and redistribution in econophysics. With this approach, we calculate the capital (wealth), the utility based on consumption, and the Gini index of these inequality using redistribution and consumption thresholds as moral parameters. The results show that: under-redistribution and waste exacerbate inequality; conversely, over-redistribution and stinginess reduce utility; and a balanced moderate moral leads to achieve both reduced inequality and increased utility. These findings provide renewed economic and numerical support for the moral importance known from philosophy, anthropology, and religion. The revival of redistribution and consumption morals should promote the transformation to a human mutual-aid economy, as indicated by philosopher and anthropologist, instead of the capitalist economy that has produced the current inequality. The practical challenge is to implement bottom-up social business, on a foothold of worker coops and platform cooperatives as a community against the state and the market, with moral consensus and its operation.

## 1. Introduction

Wealth inequality is a major social problem worldwide. According to the World Inequality Report 2022 [1], the top 10% richest people account for 76% of the world's wealth and 50% of its income, with the Gini index of income reaching 0.7. A Gini index of 0.4 is said to be a warning level for social unrest [2]. Inequality exceeding this level creates social unrest, and social unrest creates a vicious cycle of lower productivity, greater inequality, and further social unrest [3]. Therefore, solving the inequality problem is an urgent issue.

*Author for correspondence (kato.takeshi.3u@kyoto-u.ac.jp).

[1]Present address: Hitachi Kyoto University Laboratory, Kyoto University, Yoshidahonmachi, Sakyo-ku, Kyoto, 606-8501, Japan.

[2]Present address: Hitachi Central Research Laboratory, 1-280, Higashi-koigakubo, Kokubunji-shi, Tokyo 185-8601, Japan.

[3]Present address: Department of Economics, Texas Tech University, Lubbock, TX, 79409, USA.

The United Nations Sustainable Development Goals include reducing inequality as Goal 10, eradicating poverty as Goal 1, zero hunger as Goal 2, inclusive economic growth as Goal 8, and fair and inclusive institutions as Goal 16 [4]. At the World Economic Forum, a gathering of political and business leaders, the reduction of inequality is on the main agenda [5]. To achieve these goals, it is necessary to reexamine the economic relationships that create inequality.

Economist Polanyi has identified three forms of economic relations: (1) reciprocity, (2) redistribution, and (3) market exchange [6]. Philosopher Karatani presents four modes of exchange: (1) reciprocity, (2) plunder and redistribution, (3) commodity exchange, and (4) advanced recovery of reciprocity [7]. Anthropologist Graeber also presents (2) hierarchy, (3) exchange, and (4) baseline communism as three moral principles involved in economic relations [8]. To summarize, (1) is a gift economy with the obligation of return, (2) is a power economy with tax collection and redistribution, (3) is a market economy with non-human exchange of goods and money, and (4) is a human economy that sublimates the gift economy without the obligation of return. The capitalist economy that produces the current inequality is a combination of (2) a power economy and (3) a market economy [7,8], and Karatani and Graeber advocate a transformation to (4) a human economy as a prescription for the inequality problem.

Regarding economic moral, not only Graeber but also economist Bowles has stated the need to curb selfishness and foster public spirit in the face of growing inequality [9]. Philosopher Deguchi envisions a "WE society" in which free and diverse individuals "I" recognize each other's "fundamental incapability" and sublimate it into solidarity as a "WE" [10], and also proposes a "WE economy" in which individuals take on risks in accordance with their moral responsibility as coadventurers and distribute returns in consideration of their responsibility and risk vulnerability [11,12].

In "Debt: The First 5,000 Years [8]," Graeber states that human history has alternated between cycles of money and credit economies, that in the Middle Ages a moral innovation from Islam to a credit economy occurred, and that from the contemporary money economy to the next generation an innovation to a human credit economy is expected. The Islamic economy, based on the morals of the Islamic Code (*Sharia*), prohibits interest (*riba*) and speculation (*gharar*), discourages excessive consumption, and combines joint business (*mudaraba*, *murabaha*, *murabaha*, etc.) where people share profits and losses with each other and voluntary redistribution (*waqf*, *sadaqah*, and *zakat*) instead of power [13,14]. Economist Nagaoka lists the characteristics of the Islamic economy as face to face, real transactions, and mutual-aid moral [13]. The redistribution and consumption morals of the Islamic economy, as an alternative to capitalism, could be a foothold to a human economy on next generation.

Now, taking the Islamic economy as a reference, to economically examine a human economy based on moral, it is necessary to model the capital (wealth), consumption and utility of economic agents, the exchange and interaction of capital among agents, and the redistribution of capital stocked by agents. The reason for targeting consumption and utility along with capital here is that it is known that economic flow is proportional to the Gini index, i.e., economic vitalization promotes inequality [15]. There is a fear that excessive equalization and consumption restrictions will inactivate the economy and reduce the utility associated with consumption.

Macroeconomics basically deals with capital, consumption and utility of representative agents and dynamic equations for them are modelled (e.g., [16,17]). In macroeconomics, there is also an effort to introduce microeconomic heterogeneity instead of representative agents. According to the review literature in this field [18], the main topics addressed are endogenous heterogeneity of individuals and exogenous shocks, individual insurance against exogenous risks, and the interaction between idiosyncratic risks and aggregate dynamics. With respect to sources of heterogeneity, there are endogenous capabilities and preferences as initial conditions for individuals, and responses regarding transitory and persistent shocks as exogenous conditions. However, so far, morals with respect to redistribution and consumption has not been addressed. Another approach that introduces microeconomic interactions into macroeconomics is the agent-based model (ABM). According to the review literature in this field [19], macroeconomic ABM attempts to understand macro-level behaviour by having rule-based interactions among different agents [20]. However, its target agents are firms, governments, banks, and households that seek to maximize profits, and again, moral is not addressed.

In microeconomics, econophysics is an approach that focuses specifically on wealth distribution and inequality using ABM. According to the review literature in this field [21,22], wealth distribution and inequality are examined by modelling exchanges and interactions among multi agents based on physical analogies, such as the exchange of kinetic energy between two ideal gas particles. Depending on parameters of the model, such as the exchange amount, the savings rate, and the stock contribution rate, various distributions appear, including exponential, power, gamma, and delta distributions. These models have explained the basic causes of empirical distributions in the real society, e.g., Pareto's law on income and Zipf's law on city size. Recent examples of studies have examined income and inheritance taxes [23], social class and inheritance [24], tax exemption for the poor [25], contributions of surplus stock by the wealthy [15], interest/joint business and redistribution [26], and redistribution and mutual aid [27]. It should be noted that these previous studies cover the exchange and redistribution of capital, but do not deal with consumption and its associated utility.

So far, we have referred to macroeconomics and econophysics, but there is a research gap in examining the human economy, with the former regarding the morals of redistribution and consumption, and the latter regarding consumption and utility. Conversely, the former is useful in modelling capital, consumption and utility, and the latter in modelling exchange and redistribution. To examine moral, it is appropriate to introduce consumption moral into the former macroeconomic model [16,17] and to refer to the latter econophysical model of joint business and redistribution with moral [26], rather than to the anti-moral power-collecting taxation and interest business that benefit only the wealthy.

Therefore, this study aims to fill the research gap between the existing literature and the human economy, to examine the effects of redistribution and consumption morals on wealth inequality and utility, and to propose a balanced economy between both. To this end, this study applies the dynamic model of capital, consumption, and utility from macroeconomics to individual agents and the model of joint business and redistribution from econophysics to the interactions among agents. Then, by comparing wealth inequality and utility for moral differences in redistribution and consumption, we gain insight into guiding the inequality-producing capitalist economy to the human economy. Our approach is novel in that it combines a dynamic macroeconomic model with econophysical interactions and that it allows us to treat both inequality and utility simultaneously.

The remainder of this paper is organized as follows. In the Methods section, we first refer to the basic model of dynamic macroeconomics and show how to calculate the saddle point of capital and consumption and utility based on consumption with respect to individual agents. Next, following the approach of econophysics, we formulate a joint business between two agents in the contribution of capital, and its redistribution for all agents. The redistribution moral is introduced as a threshold for capital. We then show the dynamic path to the saddle point of consumption associated with changes in capital due to joint business and redistribution, and how to calculate the utility associated with changes in consumption. The consumption moral is introduced as its threshold. The Results section presents simulation results for the Gini index of wealth inequality and the utility, with redistribution and consumption thresholds as parameters. The Discussion section discusses the feasibility of a human economy based on redistribution and consumption morals as an alternative to the capitalist economy, along with an interpretation of the results. Finally, the Conclusion section presents the main conclusions and future challenges including research issues and empirical fieldwork.

# 2. Methods

## (a) Dynamic macroeconomic model of capital, consumption, and utility

First, we refer to the dynamic equations for capital and consumption in the Ramsey-Cass-Koopmans model [28–30] as the basic dynamic macroeconomic model [31,32]. If the individual agent's capital is $k$, consumption is $c$, the production function is $f(k)$, the capital depletion rate is $\delta$, the labor growth rate is $\nu$, the knowledge growth rate is $\gamma$, the discount rate of time preference is $\rho$, and the relative risk aversion coefficient is $\theta$, the dynamic equations for capital $k$ and consumption $c$ at time $t$ are expressed as equations (1) and (2).

$$\dot{k}(t) = f(k(t)) - c(t) - (\delta + \nu + \gamma)k(t). \tag{1}$$

$$\frac{\dot{c}(t)}{c(t)} = \frac{f'(k(t)) - \delta - \rho - \theta\gamma}{\theta}. \tag{2}$$

Using the Cobb-Douglas function $f(k(t)) = k(t)^\alpha$ as the production function [33,34] and setting $\nu = 0$ since the rate of labor increase is not the subject of this study, the dynamic equation can be rewritten as equations (3) and (4).

$$\dot{k}(t) = k(t)^\alpha - c(t) - (\delta + \gamma)k(t). \tag{3}$$

$$\dot{c}(t) = c(t)\frac{\alpha k(t)^{\alpha-1} - \delta - \rho - \theta\gamma}{\theta}. \tag{4}$$

The dynamic equations (3) and (4) are known to converge at the saddle point [35,36]. Solving the equations with the left-hand side of equations (3) and (4) set to zero, the capital $k^*$ and consumption $c^*$ at the saddle point are expressed as equations (5) and (6).

$$k^* = \left(\frac{\delta + \rho + \theta\gamma}{\alpha}\right)^{\frac{1}{\alpha-1}}. \tag{5}$$

$$c^* = k^{*\alpha} - (\delta + \gamma)k^*. \tag{6}$$

Next, we refer to the Ramsey-Cass-Koopmans model of utility [37,38]. The utility function $U$ is expressed as equation (7), where the instantaneous utility is $u(c(t))$, the instantaneous utility $u(c(t))$ is a constant relative risk aversion utility function $c(t)^{1-\theta}/(1-\theta)$, and assuming a finite period $t_{max}$. Here is the discount rate $\beta \equiv \rho - (1-\theta)\gamma$ in the broad sense. The meaning of this equation is the utility at $t = 0$ looking forward from $t = 0$ to $t_{max}$ in the future.

$$U(t_{max}) = \int_{t=0}^{t_{max}} e^{-\beta t}u(c(t))dt = \int_{t=0}^{t_{max}} e^{-\beta t}\frac{c(t)^{1-\theta}}{1-\theta}dt. \tag{7}$$

## (b) Econophysical model of joint business and redistribution

For joint business, we refer to the model presented in the literature [26]. In this model, $m$ pairs of two agents $i$ and $j$ ($i \neq j, i, j = 1,2,\cdots,N$) are randomly selected every period $t_{bp}$ among $N$ agents. The two agents $i$ and $j$ in each pair have capital $k_i(t)$ and $k_j(t)$, respectively, and a common savings rate $\lambda$. Both agents contribute capital, excluding savings, to the joint business, and the capital contributed by both agents, $(1 - \lambda) \cdot k_i(t)$ and $(1 - \lambda) \cdot k_j(t)$, is distributed to each agent according to the profit/loss rate $\varepsilon$ and the contribution rate. The capital $k_i(t + \Delta t)$ and $k_j(t + \Delta t)$ of the two agents $i$ and $j$ at time $t + \Delta t$ is expressed as equation (8).

$$
\begin{aligned}
k_i(t + \Delta t) &= \lambda k_i(t) + \frac{k_i(t)}{k_i(t) + k_j(t)} (1 + \varepsilon)(1 - \lambda) \left( k_i(t) + k_j(t) \right) \\
&= \lambda k_i(t) + (1 + \varepsilon)(1 - \lambda) k_i(t); \\
k_j(t + \Delta t) &= \lambda k_j(t) + \frac{k_j(t)}{k_i(t) + k_j(t)} (1 + \varepsilon)(1 - \lambda) \left( k_i(t) + k_j(t) \right) \\
&= \lambda k_j(t) + (1 + \varepsilon)(1 - \lambda) k_j(t).
\end{aligned}
\tag{8}
$$

For redistribution, we refer to the model presented in the literature [26] and set a novel threshold as the redistribution moral. In this model, each of the $N$ agents contributes capital $\mathrm{Ramp}(k_i(t) - k_{TH})$ above the threshold $k_{TH}$ for each redistribution period $t_{rp}$, and the capital $\sum_{j=1}^{N} \mathrm{Ramp}(k_j(t) - k_{TH})$ collected from the $N$ agents is redistributed to each agent according to the ratio $1/k_i(t)/\sum_{j=1}^{N} 1/k_j(t)$. That is, more capital is redistributed to the agent with less capital. The capital $k_i(t + \Delta t)$ of agent $i$ at time $t + \Delta t$ after redistribution is expressed as equations (9) and (10).

$$
k_i(t + \Delta t) = k_i(t) - \mathrm{Ramp}(k_i(t) - k_{TH}) + \frac{\frac{1}{k_i(t)}}{\sum_{j=1}^{N} \frac{1}{k_j(t)}} \sum_{j=1}^{N} \mathrm{Ramp}(k_j(t) - k_{TH}).
\tag{9}
$$

$$
\mathrm{Ramp}(x) = \begin{cases} x, \ x \geq 0. \\ 0, \ x < 0. \end{cases}
\tag{10}
$$

## (c) Adjustment of consumption and utility to changes in capital

The capital changes discontinuously before and after the joint business and redistribution. We consider that the capital of agent $i$ jumps from $k_{B_i}$ before the change to $k_{A_i}^*$ after the change, so that $k_{A_i}^*$ becomes the saddle point, and the capital depletion rate $\delta$, discount rate $\rho$ and relative risk aversion coefficient $\theta$ remain constant while the knowledge increase rate $\gamma_{A_i}$ changes. This means that the production efficiency of labor changes with the change in capital. After the change $k_{A_i}^*$ and $\gamma_{A_i}$ are expressed as equations (11) and (12), rewriting equation (5).

$$
k_{A_i}^* = \left( \frac{\delta + \rho + \theta \gamma_{A_i}}{\alpha} \right)^{\frac{1}{\alpha - 1}}.
\tag{11}
$$

$$
\gamma_{A_i} = \frac{\alpha k_{A_i}^{* \, \alpha - 1} - \delta - \rho}{\theta}.
\tag{12}
$$

When capital changes to $k_{A_i}^*$, consumption is considered to follow an adjustment path from $c_{B_i}$ before the change to the new saddle point $c_{A_i}^*$ and converge [39]. To break it down, it means that the consumption tendency does not change abruptly with the change of capital, but gradually. $c_{A_i}^*$ is expressed as equation (13), rewriting equation (6).

$$
c_{A_i}^* = k_{A_i}^{* \, \alpha} - \left( \delta + \gamma_{A_i} \right) k_{A_i}^*.
\tag{13}
$$

The change of consumption $c_{ADJ_i}(t)$ in the adjustment path is expressed as equations (14)–(17), referring to the approximation of the adjustment speed [40]. The time $t$ in these equations is the time elapsed since the time of the capital change, and $\mu_i$ is the adjustment speed (time constant).

$$c_{ADJ_i}(t) = c_{A_i}^* + e^{\mu_i t}\big(c_{B_i} - c_{A_i}^*\big). \tag{14}$$

$$\mu_i = \frac{\beta_i - \sqrt{\beta_i^2 - 4f''\big(k_{A_i}^*\big)c_{A_i}^*/\theta}}{2}. \tag{15}$$

$$\beta_i = \rho - (1-\theta)\gamma_{A_i}. \tag{16}$$

$$f''\big(k_{A_i}^*\big) = \alpha(\alpha-1)k_{A_i}^{*\ \alpha-2}. \tag{17}$$

Here, we set a novel threshold as the consumption moral. If the value of the new saddle point $c_{A_i}^*$ exceeds the threshold value $c_{TH}$, we consider that consumption will follow an adjustment path from $c_{B_i}$ before the change toward $c_{TH}$ and converge. In other words, consumption will be kept below the threshold value. The change in consumption $c_{ADJ^\#_i}(t)$ in the adjustment path is expressed as equation (18), assuming the same adjustment speed as in equation (15).

$$c_{ADJ^\#_i}(t) = c_{TH} + e^{\mu_i t}\big(c_{B_i} - c_{TH}\big). \tag{18}$$

While consumption is kept below the threshold, capital will increase. The dynamic equation of capital in this case is approximated as in equation (19) by rewriting equation (3), and the change in capital $k_{ADJ^\#_i}(t)$ is expressed as in equation (20).

$$\dot{k}_i(t) \sim k_{A_i}^{*\ \alpha} - c_{ADJ^\#_i}(t) - (\delta + \gamma_{A_i})k_{A_i}^*.$$

$$\sim k_{A_i}^{*\ \alpha} - c_{A_i}^* - (\delta + \gamma_{A_i})k_{A_i}^* + c_{A_i}^* - c_{ADJ^\#_i}(t). \tag{19}$$

$$\sim c_{A_i}^* - c_{ADJ^\#_i}(t).$$

$$k_{ADJ^\#_i}(t) \sim k_{A_i}^* + \int_0^t \Big(c_{A_i}^* - c_{ADJ^\#_i}(t)\Big)dt. \tag{20}$$

The utility $U_i$ is expressed as equation (21) with $t_{1i}, t_{2i}, \cdots, t_{ni}$ as the time when the capital changed due to the joint business or redistribution. The meaning of this equation is that the utility looking forward from the time of change to the next change occurs (to break it down, when a change occurs, the utility is reviewed again, or when time passes after the change, the instantaneous utility is discounted). The reason for this equation is that simply integrating the instantaneous utility at each point in time ($\int_{t=0}^{t_{max}} c(t)^{1-\theta}/(1-\theta)\,dt$, and if $c(t)$ is constant, $t_{max} \cdot c^{1-\theta}/(1-\theta)$) represents utility from momentary consumption without considering the future, and that equation (7) ($\int_{t=0}^{t_{max}} e^{-\beta t} c(t)^{1-\theta}/(1-\theta)\,dt$, and if $c(t)$ are constant, $1/\beta \cdot (1 - e^{-\beta t_{max}}) \cdot c^{1-\theta}/(1-\theta)$) will show the contribution of instantaneous utility becomes smaller as time $t$ approaches $t_{max}$. Note that if no change occurs at all, equation (21) is equal to equation (7). For $c_{0i}(t), c_{1i}(t), \cdots, c_{ni}(t)$ in equation (21), we can apply equation (14) or (18) depending on whether $c_{A_i}^*$ does not exceed or exceeds $c_{TH}$ at the change.

$$U_i(t_{max}) = \int_0^{t_{1i}} e^{-\beta_{0i} t}\frac{c_{0i}(t)^{1-\theta}}{1-\theta}dt + \int_{t_{1i}}^{t_{2i}} e^{-\beta_{1i}(t-t_{1i})}\frac{c_{1i}(t)^{1-\theta}}{1-\theta}dt + \cdots$$
$$+ \int_{t_{ni}}^{t_{max}} e^{-\beta_{ni}(t-t_{ni})}\frac{c_{ni}(t)^{1-\theta}}{1-\theta}dt. \tag{21}$$

## (d) Inequality and utility evaluation

Capital (wealth) inequality is evaluated by the Gini index. The Gini index is a well-known parameter for evaluating wealth inequality [41] and is calculated by equation (22) [42]. $\text{Sort}(k_i(t_{max}))$ operation sorts the capital $k_i(t_{max})$ ($i = 1, 2, \cdots, N$) of $N$ agents at time $t_{max}$ from smallest to largest, and the $q$th capital from smallest to largest is $r_q(t)$ to calculate the Gini index $g_k$. The Gini index $g_U$ of utility $U_i(t_{max})$ can be calculated in the same way.

$$r_q(t_{max}) \in \text{Sort}(k_i(t_{max})),$$

$$g_k = \frac{2 \cdot \sum_{q=1}^{N} q \cdot r_q(t_{max})}{N \cdot \sum_{q=1}^{N} r_q(t_{max})} - \frac{N+1}{N}. \tag{22}$$

The median $U_{med}$ shown in equation (23) is used as the representative value of the utility $U_i(t_{max})$ of the $N$ agents. This is because the distribution of $U_i(t_{max})$ is expected to be skewed rather than normal distribution-like. The median $k_{med}$ of the capital $k_i(t_{max})$ can be calculated in the same way.

$$s_q(t_{max}) \in \text{Sort}(U_i(t_{max})),$$

$$U_{med} = \begin{cases} s_{\frac{N+1}{2}}(t_{max}), & N \text{ is odd.} \\ \dfrac{s_{\frac{N}{2}}(t_{max}) + s_{\frac{N}{2}+1}(t_{max})}{2}, & N \text{ is even.} \end{cases} \tag{23}$$

## (e) Calculation flow

Figure 1 shows the calculation flow of capital, consumption, and utility from time $t = 0$ to $t_{max}$ (cf. code in supplementary material). First, in step 1, as initial settings, the number of agents $N$, the time increment $\Delta t$ of the calculation, the period $t_{bp}$ and the number of pairs $m$ of joint business, the period $t_{rp}$ and the start time $t_{rs}$ of redistribution, and the maximum calculation time $t_{max}$ are set. In addition, for all agents, we set the exponent $\alpha$ of the production function, the capital depletion rate $\delta$, the discount rate $\rho$, the relative risk aversion coefficient $\theta$, the initial knowledge growth rate $\gamma_0$, the savings rate $\lambda$, the variation range $\varepsilon_w$ of the profit/loss rate $\varepsilon$, and the thresholds $k_{TH}$ and $c_{TH}$ as the moral parameters for redistribution and consumption.

Step 2 calculates capital $k_0$, consumption $c_0$, and utility $U_0$ at time $t = 0$ using equations (5), (6), and (21). In step 3, advance time to $t + \Delta t$. In step 4, calculate the surplus $Mod((t_{rs} + t)/t_{rp})$ of time $t_{rs} + t$ divided by $t_{rp}$, and if it is zero, go to step 5, otherwise go to step 6. In step 5, redistribute by $k_{TH}$ using equation (9) and change $k_i(t)$ to $k_{A_i}^* = k_i(t + \Delta t)$. In step 6, the surplus $Mod(t/t_{bp})$ is calculated; if it is zero, go to step 7, otherwise go to step 8. In step 7, $m$ pairs of agents are randomly selected and conduct joint business using equation (8) to change $k_i(t)$ and $k_j(t)$ to $k_{A_i}^* = k_i(t + \Delta t)$ and $k_{A_j}^* = k_j(t + \Delta t)$ respectively.

In step 8, calculate $c_{A_i}^*$ corresponding to $k_{A_i}^*$ using equation (13), and if $c_{A_i}^* \leq c_{TH}$, go to step 9, otherwise go to step 10. In step 9, calculate $c_{ADJ_i}(t)$ as consumption $c_i(t)$ in the adjustment path according to equation (14). In step 10, calculate $c_{ADJ^{\#}_i}(t)$ and $k_{ADJ^{\#}_i}(t)$ as consumption $c_i(t)$ and capital $k_i(t)$ in the adjustment path according to equations (18) and (20), respectively. In step 11, use equation (21) to calculate the incremental $\Delta U$ of utility based on consumption $c_i(t)$ at time $t$ and utility $U_i(t)$ from time $t = 0$ to $t$.

In step 12, if $t = t_{max}$, go to step 13, otherwise return to step 3. In step 13, using equations (22) and (23), calculate the Gini index $g_k$, $g_U$, the median $k_{med}$ and $U_{med}$ for capital $k_i(t_{max})$ and utility $U_i(t_{max})$ respectively to complete the calculation for the moral parameters, threshold $k_{TH}$ and $c_{TH}$.

# 3. Results

## (a) Initial setting

First, initialize the common parameters to be used in the calculations. The number of agents is set to $N = 1,000$ (this number is sufficient to calculate the wealth distribution and the Gini index). The time increment for the calculation $\Delta t = 1/365$ (one day), the period of the joint business $t_{bp} = 1/365$ (one day) and the number of pairs $m = 17$ (on average about 1,000 people per month participate in the joint business), the period of redistribution $t_{rp} = 10$ (years) and the maximum calculation time $t_{max} = 100$ (years) (since these values are relative, for example, they can be read as 10 days, 10 months, 100 years, or 1000 years). To ensure that the timing of redistribution does not overlap with $t_{max}$ (so that the calculation results of the Gini index and utility at $t_{max}$ are not directly affected by redistribution), redistribution should begin in the fifth year ($t_{rs} = 5$).

For the exponent of the production function and the capital depletion rate in the macroeconomic model, $\alpha = 0.5$ and $\delta = 0.1$, referring to the values used in the literature [43–45]. As for the discount rate, $\rho = \ln(1/\phi) = 0.223$, converted from the value of time preference $\phi \sim 0.8$ in the literature [46]. A wide variety of values have been reported for the relative risk aversion coefficient, depending on the measurement method and subject [47–49], but here we assume $\theta = 0.5$. Assuming $\gamma_0 = 0$ for the initial rate of knowledge increase, capital and consumption are $k_0 = 2.39$ and $c_0 = 1.31$ using equations (5) and (6), respectively. The initial utility is $U_0 = 0$. These values are set as initial values for all agents.

With respect to the savings rate of joint business, $\lambda = 0.25$ with reference to the world's total savings (as a percentage of GDP) [50]. As for the profit/loss rate, it is known that the average return of stock indices is about 8%, but with large fluctuations exceeding ±10% [51], while the average return of investors is only about 2% [52]. In other words, taking into account that the business returns and losses fluctuate both positively and negatively, we set uniform random numbers in the range $-\varepsilon_w \leq \varepsilon \leq \varepsilon_w$ ($\varepsilon_w = 0.1$).

With the above initial settings, the Gini index and median values of capital and utility are calculated according to the calculation flow in figure 1, using the redistribution threshold $k_{TH}$ and the consumption threshold $c_{TH}$ as moral parameters.

## (b) Calculation results

Figures 2a1–2a3 show the calculation results for the redistribution threshold $k_{TH} = 100$ and the consumption threshold $c_{TH} = 100$ (low moral), and figures 2b1–2b3 for $k_{TH} = 1.7$ and $c_{TH} = 5.5$ (high moral). 2a1 and 2b1 are histograms of capital $k(t)$, 2a2 and 2b2 of consumption $c(t)$, and 2a3 and 2b3 of utility $U(t)$ at times $t = 1$, 30 and 100 ($t_{max}$). The distributions of capital $k(t)$, consumption $c(t)$, and utility $U(t)$ change over time. The distribution of $U(t)$ shifts to the right over time because instantaneous utility is integrated over time. Comparing figures 2a1–2a3 and 2b1–2b3, we can see that when the morals of redistribution and consumption are low, the distributions of capital, consumption, and utility gradually widen and become exponential-like distribution at $t_{max}$ (inequality increases). In the case of high moral, the distributions of capital, consumption, and utility do not widen and become like a normal-like distribution at $t_{max}$ (inequality is suppressed). Note that in the literature [26,27], it is known that if redistribution is not done at all, the distribution will eventually become delta distribution over time (the Gini index will be 1).

Figure 3 shows time series graphs from time $t = 1$ to $100$ ($t_{max}$) for three representative agents with $k_{TH} = c_{TH} = 100$ and $k_{TH} = 1.7, c_{TH} = 5.5$. *3a1* and *3b1* are capital $k(t)$, *3a2* and *3b2* are consumption $c(t)$, *3a3* and *3b3* are utilities $U(t)$. Capital $k(t)$ changes discontinuously through joint business and redistribution, while consumption $c(t)$ follows changes in capital with an adjustment speed (time constant). Utility $U(t)$ integrates the instantaneous utility based on consumption, so its change is gradual and shows only a slight inflection. Comparing figures *3a1–3a3* and *3b1–3b3*, we see that when redistribution and consumption morals are low, the differences in capital, consumption, and utility widen among agents (corresponding to the distribution in figures *2a1–2a3*), and that when morals are high, these differences do not widen and remain within a certain range (corresponding to the distribution in figures *2b1–2b3*). We also see that when morals are high, the amount of capital and consumption among agents interchange with each other, i.e., inequality is not fixed.

Figure 4 is a three-dimensional graph with the redistribution threshold $k_{TH}$ on the x-axis, the consumption threshold $c_{TH}$ on the y-axis, capital $k_{med}$, utility $U_{med}$, Gini index of capital $g_k$, and Gini index of utility $g_U$ at time $t = t_{max}$ on the z-axes (see data in supplementary material). Figure *4a* shows that as $c_{TH}$ is made smaller (the higher the consumption moral), the larger the capital, but if $k_{TH}$ is too large (the lower the redistribution moral), the smaller the capital. In other words, both redistribution and consumption morals are necessary. *4b* shows that the smaller $k_{TH}$ (the higher the redistribution moral), the greater the utility, but if $c_{TH}$ is too small (too stingy), the smaller the utility. In other words, it is better to be moderately thrifty without being stingy in consumption moral. *4c* shows that the smaller $k_{TH}$ is (the higher the redistribution moral), the more capital inequality is suppressed, but that making $k_{TH}$ and $c_{TH}$ too small will produce inequality conversely. This is because if all agents are stingy and even those with small capital redistribute, the capital of the poor will be cut even more, and the distribution of capital will not be narrowed. In other words, there is no need to impose the moral of redistribution on the poor. *4d* shows that smaller $k_{TH}$ and $c_{TH}$ (the higher the redistribution and consumption morals) reduce utility inequality. To summarize what we can see from figure 4, moderate and balanced redistribution and consumption morals are important, not extreme. Table 1 summarizes the main effects of the redistribution threshold $k_{TH}$ and the consumption threshold $c_{TH}$ on capital and utility.

In figure 5, we introduce a novel balance index between inequality and utility to look at the balance between redistribution and consumption morals (see data in supplementary material). Figure *5a* is a three-dimensional graph with the redistribution threshold $k_{TH}$ on the x-axis, the consumption threshold $c_{TH}$ on the y-axis, and the balance index $U_{med}/g_k$, that is utility $U_{med}$ divided by the Gini index $g_k$ of capital, on the z-axis, and *5b* is a contour graph of *5a*. The shape of both graphs shows that for $k_{TH}$, a steep slope is observed in the region where $k_{TH}$ is small, while for $c_{TH}$, a gentle slope is observed. *5b* has a $U_{med}/g_k$ peak (cross mark in the figure) at approximately $k_{TH} \sim 1.7$ and $c_{TH} \sim 5.5$. The fact that it is mountainous indicates that balancing redistribution and consumption morals can both reduce inequality and increase utility. The slowness with respect to $c_{TH}$ is due to the fact that utility depends on the time constant and integral of consumption. Note that compared to the initial values of $k_0 = 2.39$ and $c_0 = 1.31$ for all agents, the peak $k_{TH}$ value is smaller and the $c_{TH}$ value is much larger. This implies that it is important to raise the redistribution moral to reduce and save the inequality that biases the distribution toward the poor, as in an exponential distribution, while the consumption moral does not need to be so strict as long as it is not extremely wasteful. Another way of looking at it is that if redistribution is done well enough, consumption moral can be relaxed because the saddle point of consumption is seldom too large.

Approximating the graph in figure 5 using a Gauss-type function $e^{-(x-a)^2}$ with $x$ as a variable and $a$ as a peak, and the peak values $k_{TH} \sim 1.7$ ($\ln 1.7 = 0.53$) and $c_{TH} \sim 5.5$ ($\ln 5.5 = 1.7$), the graph is expressed as equation (24). The coefficient of determination is $R^2 = 0.97$, which is a highly accurate approximation.

$$\frac{U_{med}}{g_k} \sim 390 \, e^{-(\ln k_{TH} - 0.53)^2 - 0.037 \, (\ln c_{TH} - 1.7)^2} + 361. \tag{24}$$

In the lower left part of figure 5*b*, the mountain shape (contour lines) extends to the lower right. The dashed line in the figure is expressed as equation (25). This equation indicates that a larger $k_{TH}$ requires a smaller $c_{TH}$, and conversely, a larger $c_{TH}$ requires a smaller $k_{TH}$, i.e., the morals of redistribution and consumption are complementary to each other.

$$k_{TH} \times c_{TH} \sim 5. \tag{25}$$

Figure 6 shows a correlation graph with the Gini index of capital, $g_k$, on the x-axis and utility, $U_{med}$, on the y-axis. A negative correlation is observed where the larger $g_k$, the smaller $U_{med}$. This indicates that reducing inequality leads to higher utility. The approximation is $y = 239 - 87x$, and the p-value is less than 0.01.

## 4. Discussion

The calculation results of the capital, consumption, and utility for joint business and redistribution reveal that redistribution and consumption morals are effective in reducing inequality and increasing utility, and that there is a moderate region that balances both morals. These things were known empirically, for example, in the discourses of philosophers and anthropologists and as religious mores, but the novel approach coupling the dynamic macroeconomic model with the econophysical model has anew supported them economically and numerically. Furthermore, the novel introduction of inequality and utility balance index reveals that the balance region between redistribution and consumption is represented by a Gauss-type peak function. These results reaffirm the importance of moral in the economy.

According to a study on redistribution in the literature [27], it is known that the relationship $(f/g)$ between economic flows $(f)$ and the Gini index $(g)$ can be approximated by a Box Lucas-type function $a(1 - e^{-bx})$ with the redistribution transfer rate $x$ as a variable. Both this Box Lucas type and the Gauss type $e^{-(x-a)^2}$ of this study are based on exponential functions, but the former is a saturated function and the latter is a peak function. The difference between them can be attributed to the difference in whether the redistribution method is a "ratio" or a "threshold" of all agents to capital, and whether the numerator of the objective function is economic flows (the amount of capital in circulation) or utility. Future research comparing variations in redistribution methods and targeting both economic flows and utility could provide comprehensive findings that link both.

In this study, the threshold was set as the moral of consumption as well as the moral of redistribution, but it is also possible to set a threshold for the discount rate (time preference). In this case, we would be modelling constraints on future consumption rather than constraints on current consumption. The basic trend may not change between this model and the current model, but the temporal nature of consumption moral will be examined. In addition, more detailed findings could be obtained by considering the heterogeneity of agents' discount rates (e.g., [53]) and production functions, as well as the effect of consumption restraint on the profit/loss rate. In the future, instead of the Ramsey-Cass-Koopmans model, there may be comparisons with other functional forms of utility functions (e.g., [54]) and development into a dynamic stochastic general equilibrium model [55].

The present results showed a negative correlation between utility and the Gini index, and it is generally known that there is a negative correlation between life satisfaction and the Gini index [56]. Although life satisfaction does not depend solely on utility for consumption, the present results support the influence of the Gini index. According to a global life satisfaction survey [57], the distribution of life satisfaction on a scale of 0 to 10 is skewed toward the side greater than 5 in Western Europe, followed by Central and Eastern Europe, North America, and Latin America, symmetrically distributed around 5 in East Asia, Southeast Asia, Middle East and North Africa, and skewed toward the side of distribution smaller than 5 in South Asia and sub-Saharan Africa. Although it is difficult to generalize due to cultural differences, the high level of life satisfaction in Western Europe may be related to the fact that social security, i.e., redistribution, is more extensive. However, this is (2) tax collection and redistribution by power, one of the economic modes mentioned in the Introduction section, and it is necessary to change this to (4) human economy based on moral as described by Karatani and Graeber [7,8].

Simply put, the morals of redistribution and consumption are "knowing how much is enough." For example, the Islamic Quran states "Those who spend neither wastefully nor stingily, but moderately in between,"and teaches a moderation, i.e., thrift, between waste and stinginess [58]. In addition, Lao Tzu said "Those who know when they have enough are rich [59]," Buddhism has "Desiring less and being satisfied with what I already have [60]," Hindu yoga has "contentment and a lack of desire for what others have [61]," and the Christian Bible says "The sleep of a laborer is sweet, whether they eat little or much, but as for the rich, their abundance permits them no sleep [62]." At the United Nations, the "Sufficiency Economy" advocated by King Bhumibol of Thailand [63] is attracting attention.

While a moral revival is awaited for the transformation to a human economy described by Karatani and Graeber, there is something to keep in mind. Deguchi proposes a mutual-aid "WE society" [10], while expressing concern that moral enforcement will lead to an oppressive, exclusive, and totalitarian society [64]. He also states that rather than imposing morals and values and excluding those who do not conform to them, softening WE by breaking the restraints will lead to wellbeing. Philosopher Heath states that social norms arise from mutual expectations and approval [65], and economist Aoki states that self-sustaining norms arise from shared group expectations [66]. Therefore, morals will not be enforced, but will need to be fostered bottom-up in the community, or in Islam, in *ummah*.

Sociologist Pestoff, economist Rajan, and policy scholar Hiroi have identified the community as a third pole to confront with the state and the market, describing a balance and synthesis of the three [67–69]. If (2) tax collection and redistribution by power corresponds to the state and (3) exchange of goods and money corresponds to the market among the economic modes, then (4) a human economy based on moral, while grounded in community, would replace the role of the state with the moral of voluntary redistribution and reflect that moral for exchange and consumption in the market. Economist Ostrom cites the arrangement of operating rules through collective choice as one of the design principles of the commons [70]. In other words, this is the fostering of morals through consensus building. For example, in modern Japanese mutual-aid communities, as a moral practice, half of the revenue from natural resources was used for current and temporary expenditures, and the other half was used for other transfers (donations and charity) and self-giving (future reserves) [71].

In light of the above, the transformation from a capitalist economy that generates inequality to a human economy based on moral will begin at the grassroots level, starting with communities. Specifically, these include worker co-ops [72], where workers combine labor, investment, and management; platform cooperatives [73,74], where users and workers engage in joint ownership and democratic decision-making; and community wealth building [75], which supports asset ownership and employment in local communities. Implementing redistribution and consumption morals as self-governance for running these social businesses will lead to an economy that balances inequality and utility. Furthermore, in the future, it will be useful to utilize information systems such as the Social Co-Operating System, which combines an operational loop that promotes morally based behavior and a collegial loop that supports moral consensus building [76].

# 5. Conclusions

In this study, we have evaluated the effects of redistribution and consumption morals on wealth inequality and utility using a novel approach that couples a dynamic macroeconomic model with an econophysical model. These morals were set as thresholds for redistribution and consumption, respectively, and the Gini index of inequality and consumption-based utility were evaluated. The new findings are that: (1) redistribution moral is mainly effective in reducing capital and utility inequality and increasing utility; (2) consumption moral is mainly effective in increasing capital and reducing utility inequality; (3) striking a moderate balance between redistribution and consumption morals achieve both reducing inequality and increasing utility; and (4) redistribution and consumption morals are complementary but the former is more effective. These findings provide a new grasp of the importance of moral, which has been discussed in philosophy, anthropology, and religion, from an economic perspective.

Although this study used only basic macroeconomic and econophysics models, its basic nature allowed us to discard details and focus on morals to clarify their effectiveness, and to suggest a direction for change from an inequality-producing capitalist economy to a human economy. Future issues include: variations in modelling redistribution and consumption morals, introducing agent heterogeneity and moral temporality, expanding to economic flows in addition to inequality and utility as evaluation indices, case studies and field empirical studies of moral-based economic activity, and social movements from a capitalist economy to a human economy that corrects inequality.

## Acknowledgments

We are deeply grateful to Prof. Yasuo Deguchi of Kyoto University and the Kyoto Institute of Philosophy, and Emeritus Prof. Yoshinori Hiroi of Kyoto University, for their useful suggestions on "WE society" and post-capitalism, respectively. Furthermore, we would like to extend our gratitude to our colleagues at the Hitachi Kyoto University Laboratory of Kyoto University for their ongoing cooperation.

# Tables

**Table 1.** Primary effects of the redistribution threshold $k_{TH}$ and the consumption threshold $c_{TH}$ on capital median $k_{med}$, Gini index $g_k$, utility median $U_{med}$, and Gini index $g_U$.

| effected variable | | smaller redistribution threshold $k_{TH}$ (higher moral) | smaller consumption threshold $c_{TH}$ (higher moral) |
|---|---|---|---|
| capital (wealth) | median $k_{med}$ | — | increase wealth |
| | Gini index $g_k$ | reduce wealth inequality | — |
| utility | median $U_{med}$ | increase utility | — |
| | Gini index $g_U$ | reduce utility inequality | reduce utility inequality |

# Figures

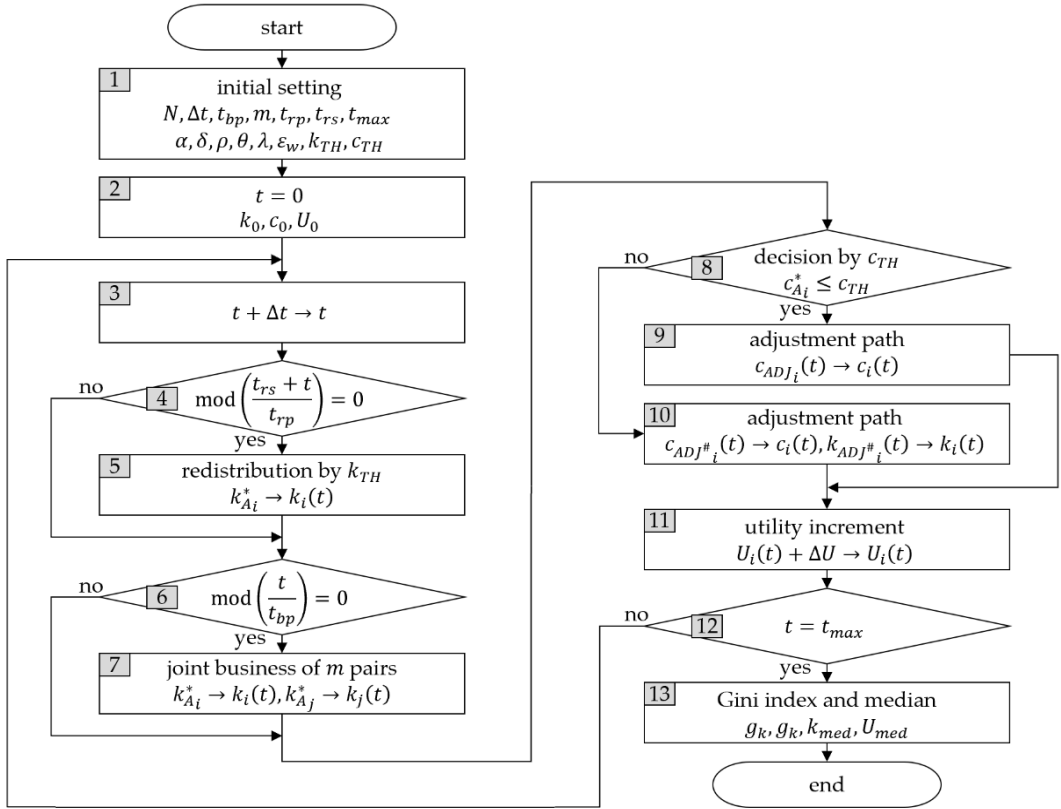

**Figure 1.** Calculation flow of capital, consumption, and utility for the redistribution threshold $k_{TH}$ and the consumption threshold $c_{TH}$ as moral parameters.

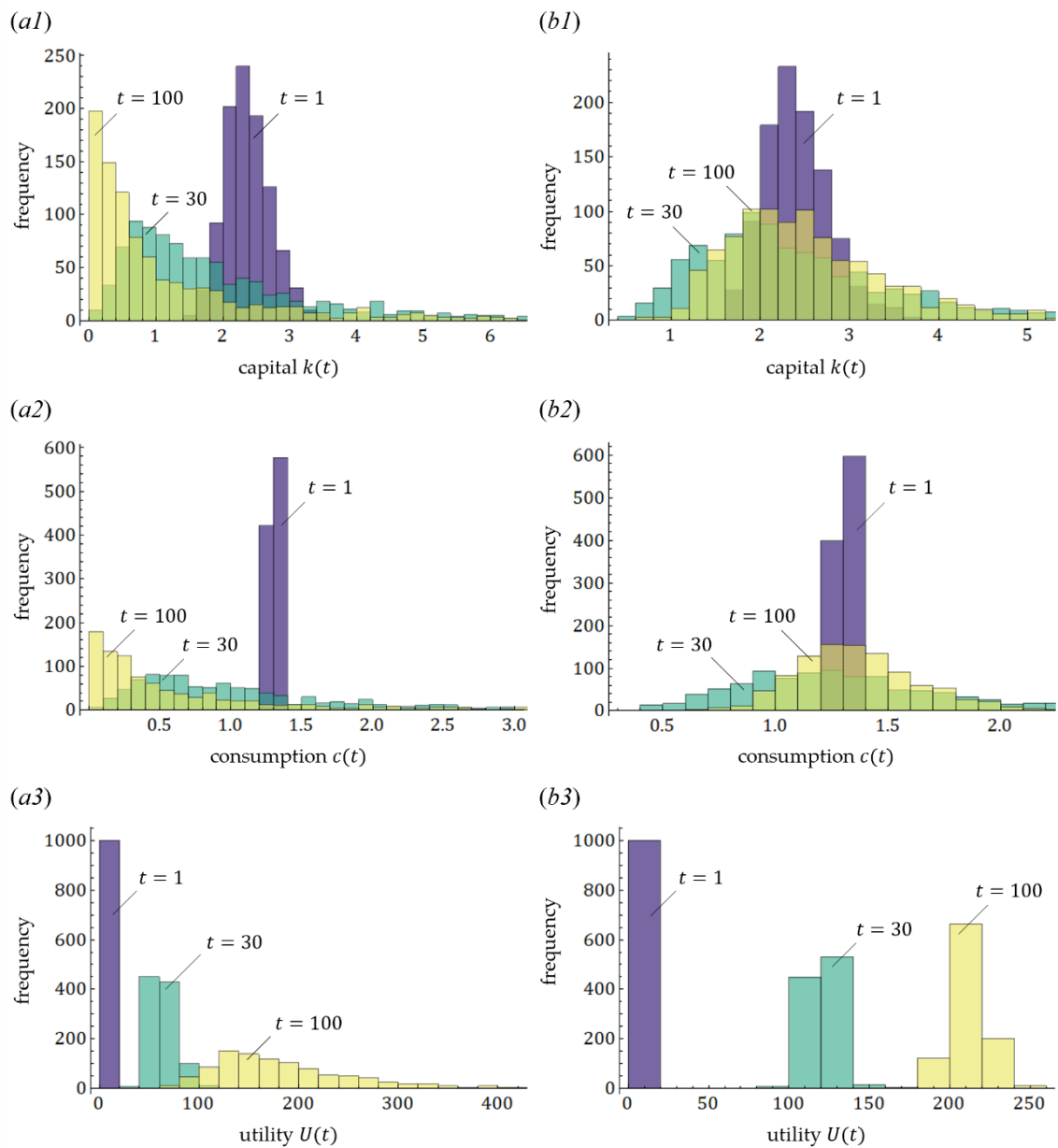

**Figure 2.** Histograms for (*a1*)–(*a4*) the redistribution and consumption thresholds $k_{TH} = c_{TH} = 100$ and (*b1*)–(*b4*) $k_{TH} = 1.7, c_{TH} = 5.5$ at $t = 1$, 30 and 100 ($t_{max}$). (*a1*) and (*b1*) for capital $k(t)$, (*a2*) and (*b2*) for consumption $c(t)$, and (*a3*) and (*b3*) for utility $U(t)$.

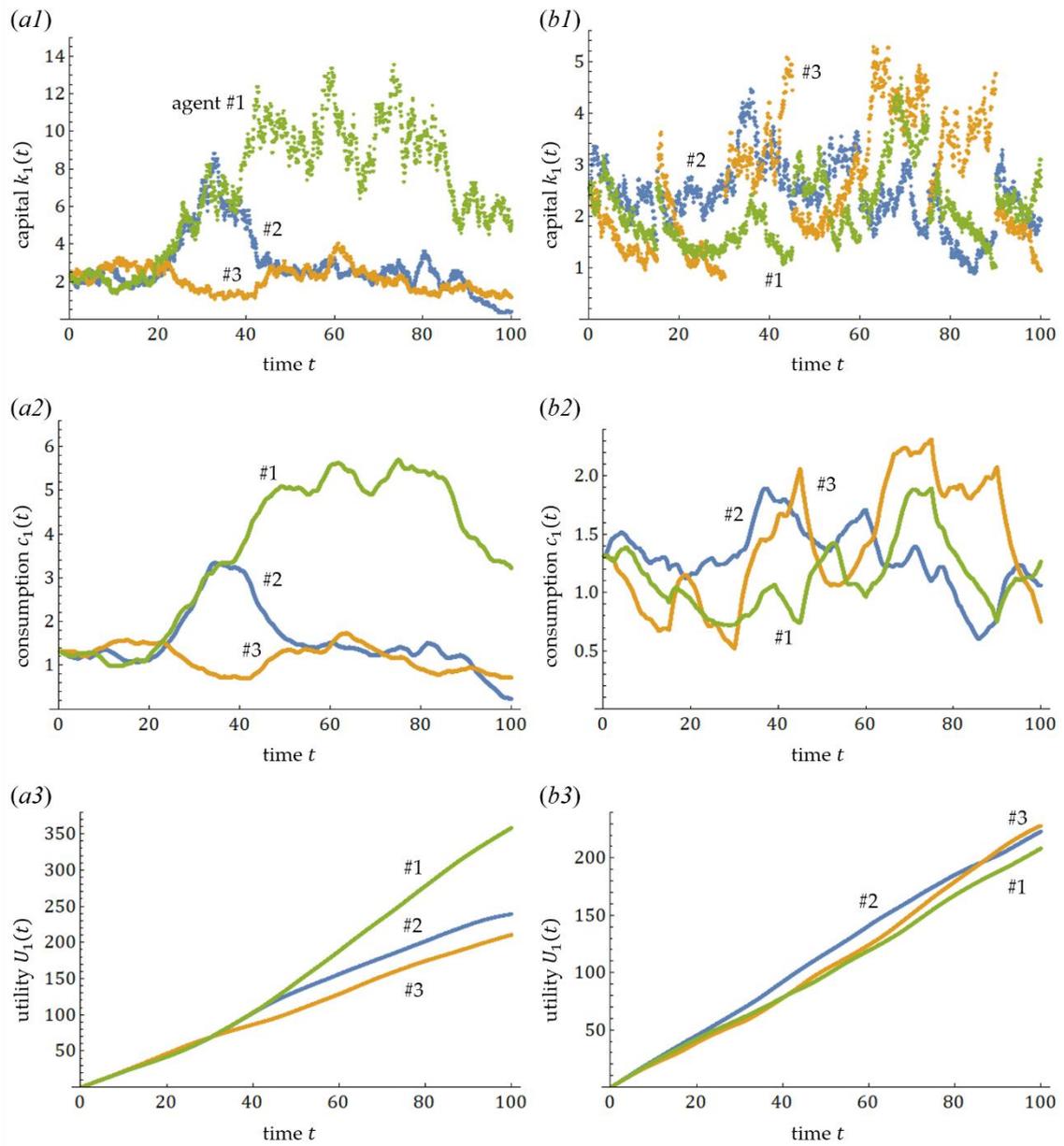

**Figure 3.** Time series graphs of three agents for (*a1*)–(*a4*) the redistribution and consumption thresholds $k_{TH} = c_{TH} = 100$ and (*b1*)–(*b4*) $k_{TH} = 1.7, c_{TH} = 5.5$ at $t = 1$, 30 and 100 ($t_{max}$). (*a1*) and (*b1*) for capital $k(t)$, (*a2*) and (*b2*) for consumption $c(t)$, and (*a3*) and (*b3*) for utility $U(t)$.

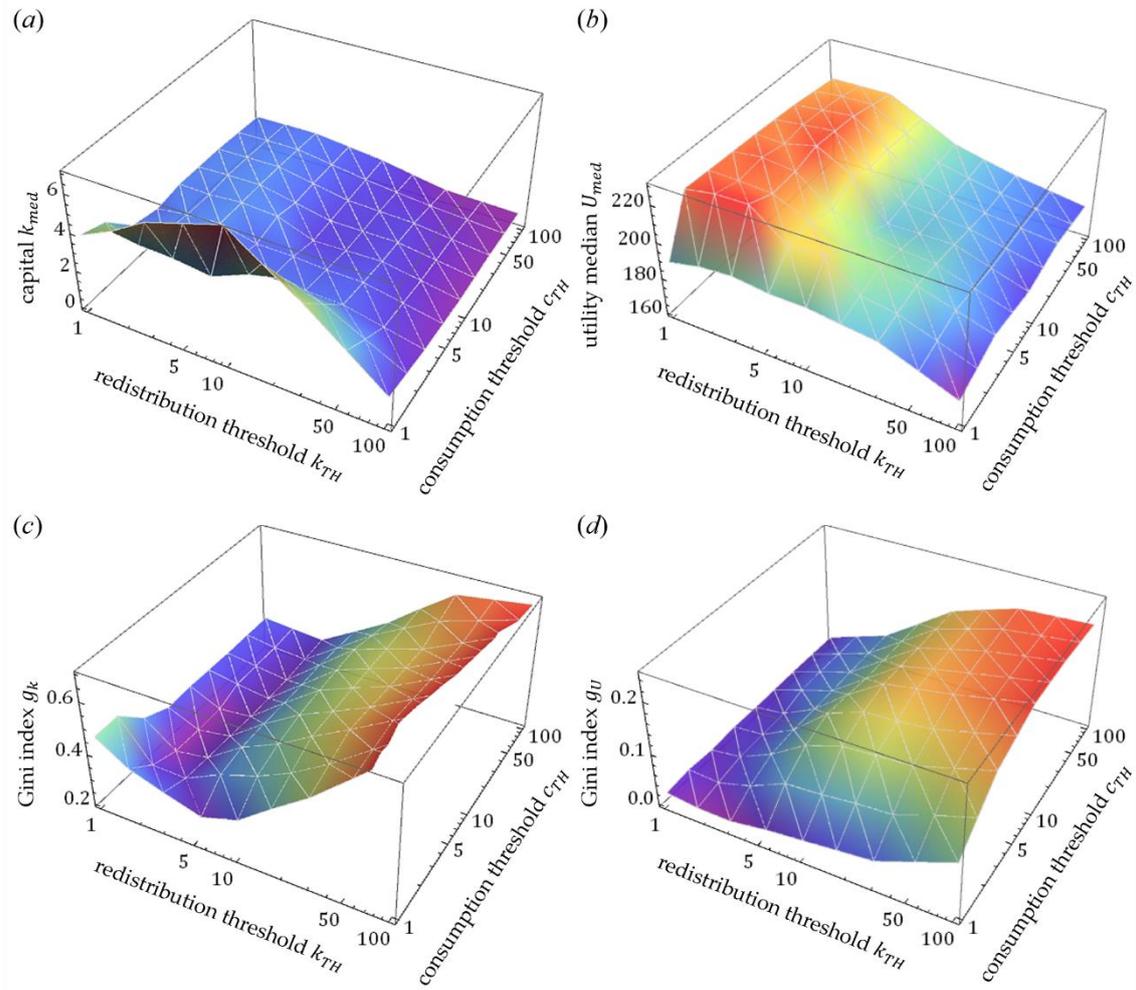

**Figure 4.** Three dimensional graphs for (*a*) capital median $k_{med}$, (*b*) utility median $U_{med}$, (*c*) Gini index $g_k$ of capital, and (*d*) Gini index $g_U$ of utility as z-axes with the redistribution threshold $k_{TH}$ as x-axes and the consumption threshold $c_{TH}$ as y-axes at $t = t_{max}$.

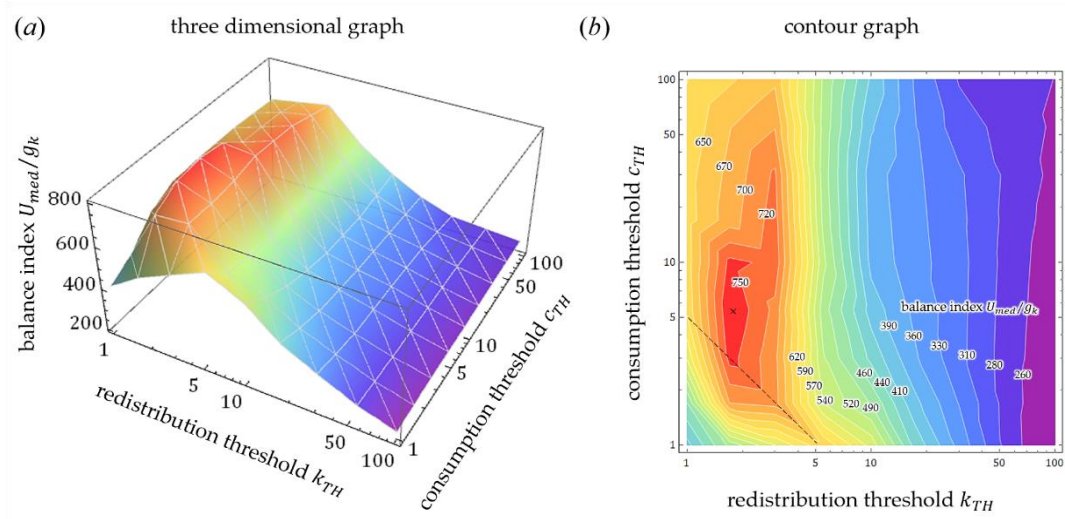

**Figure 5.** (*a*) Three dimensional graph and (*b*) contour graph for the balance index $U_{med}/g_k$ of inequality and utility as z-axis and contour respectively, with the redistribution threshold $k_{TH}$ as x-axes and the consumption threshold $c_{TH}$ as y-axes.

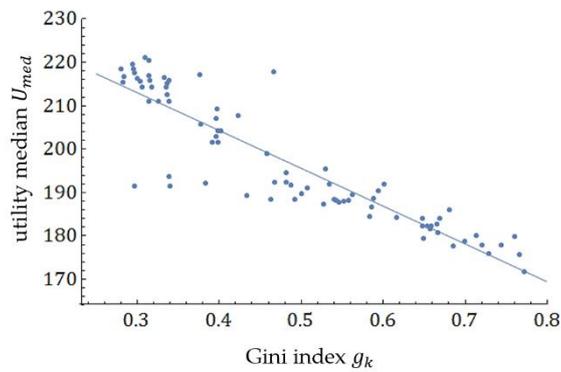

**Figure 6.** Correlation graph with Gini index $g_k$ of capital as x-axis and utility median $U_{med}$ as y-axis.

# Supplementary material

# Code

Example code to calculate capital median $k_{med}$, utility median $U_{med}$, Gini index $g_k$ of capital, Gini index $g_U$ of utility, and balance index $U_{med}/g_k$ of inequality and utility at $t = t_{max}$ for the redistribution threshold $k_{TH}$ and the consumption threshold $c_{TH}$ as moral parameters.

```
(*Initial setting------------------------------------------------------------------------------------------------*)
n=1000(*number of agents*);
Δt=1/365(*time increment,1 day*);tmax=100(*maximum calculation time, 100 years*);

(*Ramsey-Cass-Koopmans model*)
α=0.5(*exponent of production function*);δ=0.1(*capital depletion rate*);
θ=0.5(*relative risk aversion coefficient*);
ϕ=0.8(*time preference*);ϱ=Log[1/ϕ](*discount rate*);
γ0=0(*initial knowledge increase rate*);
k0=((δ + ϱ + θ * γθ)/α)^(1/(α-1))          (*initial capital*);c0=k0^α-(δ+γ0)*k0(*initial consumption*);
β0=ϱ-(1-θ)*γ0(*variable of adjustment speed*);
ff0=α*(α-1)*k0^(α-2)(*variable of adjustment speed*);
u0=0(*initial instantaneous utility*);

(*redistribution model*)
trp=10(*period, 10 years*);

(*joint business model*)
λ=0.25(*saving rate*);ςw=0.1(*width of profit/loss rate*);
tbp=1/365(*period, 1 day*);m=17(*number of pairs*);

(*initial value of agent at t=0*)
k[0]=Table[k0,{i,1,n}];c[0]=Table[c0,{i,1,n}];(*capital and consumption*)
kb[0]=Table[k0,{i,1,n}];cb[0]=Table[c0,{i,1,n}];(*before change in capital*)
ka[0]=Table[k0,{i,1,n}];ca[0]=Table[c0,{i,1,n}];(*after change in capital*)
kaa[0]=Table[k0,{i,1,n}];(*after change in capital, consumption exceeds threshold*)
γa[0]=Table[γ0,{i,1,n}];(*knowledge increase rate*)
u[0]=Table[u0,{i,1,n}];(*instantaneous utility*)
μ[0]=Table[0,{i,1,n}];tadj[0]=Table[0,{i,1,n}];(*speed and time in adjustment path*)
β[0]=Table[β0,{i,1,n}];ff[0]=Table[ff0,{i,1,n}];(*variable of adjustment speed*)
ii[0]=0;jj[0]=0;(*agent number*)

(*moral parameter-------------------------------------------------------------------------------------------*)
kth=100(*redistribution threshold*);cth=100(*consumption threshold*);
```

```
(*calculation from t=Δt to tmax--------------------------------------------------------------------------*)
Do[

(*setting at t*)
k[t]=k[t-Δt];c[t]=c[t-Δt];
kb[t]=kb[t-Δt];cb[t]=cb[t-Δt];
ka[t]=ka[t-Δt];ca[t]=ca[t-Δt];
kaa[t]=kaa[t-Δt];
γa[t]=γa[t-Δt];
u[t]=u[t-Δt];
μ[t]=μ[t-Δt];tadj[t]=tadj[t-Δt];
β[t]=β[t-Δt];ff[t]=ff[t-Δt];
ii[t]=ii[t-Δt];jj[t]=jj[t-Δt];

(*redistribution, new saddle point of capital--------------------------*)
If[Mod[t,trp+5]<=0,
(*redistribution of capital*)
ka[t]=Table[k[t][[i]]-Ramp[k[t][[i]]-kth]+1/k[t][[i]]/Sum[1/k[t][[j]],{j,1,n}]*Sum[Ramp[k[t][[j]]-kth],{j,1,n}],{i,1,n}];
(*conserving old consumption*)
cb[t]=Table[c[t][[i]],{i,1,n}];
(*time origin of adjustment*)
tadj[t]=Table[t,{i,1,n}]];

(*joint business, new saddle point of capital--------------------------*)
If[Mod[t,tbp]<=0,
Do[
(*selection of agent pair*)
ii[t]=RandomInteger[{1,n}];jj[t]=RandomInteger[{1,n-1}];
If[ii[t]<=jj[t],jj[t]=jj[t]+1,jj[t]=jj[t]];
(*setting of profit/loss ratio*)
ϵ=RandomReal[{-ϵw,ϵw}];
(*distribution of capital*)
ka[t]=ReplacePart[ka[t],
    {ii[t]->λ[t][[ii[t]]]*k[t][[ii[t]]]+(1+ϵ)*(1-λ[t][[ii[t]]])*k[t][[ii[t]]],
    jj[t]->λ[t][[jj[t]]]*k[t][[jj[t]]]+(1+ϵ)*(1-λ[t][[jj[t]]])*k[t][[jj[t]]]}];
(*conserving old consumption*)
cb[t]=ReplacePart[cb[t],{ii[t]->c[t][[ii[t]]],jj[t]->c[t][[jj[t]]]}];
(*time origin of adjustment*)
tadj[t]=ReplacePart[tadj[t],{ii[t]->t,jj[t]->t}]
,m];(*number of pairs*)

(*adjustment to change in capital--------------------------------------------*)
(*new saddle point of consumption*)
γa[t]=Table[α*ka[t][[i]]^(α-1)-δ-ϱ)/θ,{i,1,n}];
ca[t]=Table[ka[t][[i]]^α-(δ+γa[t][[i]])*ka[t][[i]],{i,1,n}];
```

```
β[t]=Table[ϱ-(1-θ)*γa[t][[i]],{i,1,n}];
ff[t]=Table[α*(α-1)*ka[t][[i]]^(α-2),{i,1,n}];

(*Determination of saddle point of consumption*)
Do[If[ca[t][[i]]<=cth,
(*not exceed threshold*)
μ[t]=ReplacePart[μ[t],{i->(β[t][[i]]-√(β[t][[i]]²-4*ff[t][[i]]*ca[t][[i]]/θ))/2}];
k[t]=ReplacePart[k[t],{i->ka[t][[i]]}];
c[t]=ReplacePart[c[t],{i->ca[t][[i]]+Exp[μ[t][[i]]*(t-tadj[t][[i]])]*(cb[t][[i]]-ca[t][[i]])}],
(*exceed threshold*)
μ[t]=ReplacePart[μ[t],{i->(β[t][[i]]-√(β[t][[i]]²-4*ff[t][[i]]*ca[t][[i]]/θ))/2}];
c[t]=ReplacePart[c[t],{i->cth+Exp[μ[t][[i]]*(t-tadj[t][[i]])]*(cb[t][[i]]-cth)}];
k[t]=ReplacePart[k[t],{i->ka[t][[i]]+(ca[t][[i]]-c[t][[i]])*Δt}]]
,{i,1,n}];

(*utility increment----------------------------------------------------------*)
Do[u[t]=ReplacePart[u[t],
    {i->u[t][[i]]+Exp[-β[t][[i]]*(t-tadj[t][[i]])]*c[t][[i]]^(1-θ)/(1-θ)*Δt]}
,{i,1,n}]

,{t,Δt,tmax,Δt}]
(*end of calculation loop ---------------------------------------------------------------------*)

(*median---------------------------------------------------------------*)
(*capital*)
kmed=Median[k[tmax]]
(*utility*)
umed=Median[u[tmax]]

(*Gini index--------------------------------------------------------*)
(*capital*)
gk=Sort[k[tmax]];ginik=(2*Sum[i*gk[[i]],{i,1,n}])/(n*Sum[gk[[i]],{i,1,n}])-(n+1)/n
(*utility*)
gu=Sort[u[tmax]];giniu=(2*Sum[i*gu[[i]],{i,1,n}])/(n*Sum[gu[[i]],{i,1,n}])-(n+1)/n

(*end-------------------------------------------------------------------------------------------*)
```

# Data

**Table S1.** Calculation results of capital median $k_{med}$, utility median $U_{med}$, Gini index $g_k$ of capital, Gini index $g_U$ of utility, and balance index $U_{med}/g_k$ of inequality and utility at $t = t_{max}$ for the redistribution threshold $k_{TH}$ and the consumption threshold $c_{TH}$ as moral parameters.

| $k_{TH}$ | $c_{TH}$ | $k_{med}$ | $U_{med}$ | $g_k$ | $g_U$ | $U_{med}/g_k$ |
|---|---|---|---|---|---|---|
| 1 | 1 | 3.847 | 188.5 | 0.4634 | 0.02015 | 406.8 |
| 1 | 1.7 | 3.430 | 217.8 | 0.4659 | 0.02512 | 467.4 |
| 1 | 3 | 2.089 | 217.1 | 0.3763 | 0.02201 | 577.0 |
| 1 | 5.5 | 1.800 | 210.2 | 0.3335 | 0.02569 | 630.3 |
| 1 | 10 | 2.041 | 215.9 | 0.3391 | 0.02472 | 636.8 |
| 1 | 17 | 2.038 | 215.1 | 0.3370 | 0.02541 | 638.4 |
| 1 | 30 | 2.035 | 214.3 | 0.3348 | 0.02615 | 640.1 |
| 1 | 55 | 1.948 | 212.6 | 0.3367 | 0.02555 | 631.6 |
| 1 | 100 | 1.863 | 210.9 | 0.3385 | 0.02497 | 623.2 |
| 1.7 | 1 | 4.629 | 192.3 | 0.3837 | 0.01254 | 501.1 |
| 1.7 | 1.7 | 2.815 | 220.5 | 0.3144 | 0.02122 | 701.5 |
| 1.7 | 3 | 2.262 | 216.8 | 0.2835 | 0.03010 | 764.6 |
| 1.7 | 5.5 | 2.039 | 218.4 | 0.2805 | 0.03073 | 778.8 |
| 1.7 | 10 | 1.991 | 215.3 | 0.2826 | 0.03167 | 761.9 |
| 1.7 | 17 | 2.093 | 217.0 | 0.3146 | 0.04441 | 689.7 |
| 1.7 | 30 | 2.061 | 215.9 | 0.3159 | 0.04560 | 683.6 |
| 1.7 | 55 | 1.975 | 214.2 | 0.3177 | 0.04501 | 674.4 |
| 1.7 | 100 | 1.878 | 210.9 | 0.3264 | 0.04580 | 646.2 |
| 3 | 1 | 5.718 | 193.8 | 0.3393 | 0.01129 | 571.2 |
| 3 | 1.7 | 2.755 | 221.1 | 0.3101 | 0.03919 | 712.9 |
| 3 | 3 | 2.098 | 215.7 | 0.3039 | 0.05891 | 709.9 |
| 3 | 5.5 | 2.175 | 216.3 | 0.3001 | 0.06311 | 720.9 |
| 3 | 10 | 2.151 | 219.6 | 0.2943 | 0.06268 | 746.2 |
| 3 | 17 | 2.120 | 218.6 | 0.2955 | 0.06383 | 739.7 |
| 3 | 30 | 2.088 | 217.5 | 0.2969 | 0.06505 | 732.7 |
| 3 | 55 | 1.990 | 214.2 | 0.3057 | 0.06585 | 700.7 |
| 3 | 100 | 1.894 | 211.0 | 0.3144 | 0.06664 | 670.9 |
| 5.5 | 1 | 4.649 | 191.5 | 0.2972 | 0.02250 | 644.4 |
| 5.5 | 1.7 | 2.521 | 205.8 | 0.3773 | 0.07126 | 545.4 |
| 5.5 | 3 | 2.066 | 207.1 | 0.3962 | 0.10193 | 522.7 |
| 5.5 | 5.5 | 1.727 | 201.5 | 0.3987 | 0.11958 | 505.5 |
| 5.5 | 10 | 2.029 | 209.4 | 0.3972 | 0.11566 | 527.0 |
| 5.5 | 17 | 1.739 | 204.1 | 0.3979 | 0.11083 | 513.0 |
| 5.5 | 30 | 1.741 | 204.3 | 0.4024 | 0.11100 | 507.7 |
| 5.5 | 55 | 1.710 | 203.0 | 0.3967 | 0.11140 | 511.7 |
| 5.5 | 100 | 1.678 | 201.6 | 0.3909 | 0.11181 | 515.8 |

**Table S1.** (continued)

| $k_{TH}$ | $c_{TH}$ | $k_{med}$ | $U_{med}$ | $g_k$ | $g_U$ | $U_{med}/g_k$ |
|---|---|---|---|---|---|---|
| 10 | 1 | 6.931 | 191.4 | 0.3400 | 0.02963 | 563.1 |
| 10 | 1.7 | 3.352 | 207.7 | 0.4228 | 0.07575 | 491.2 |
| 10 | 3 | 1.902 | 199.0 | 0.4580 | 0.12477 | 434.5 |
| 10 | 5.5 | 1.713 | 194.6 | 0.4810 | 0.14648 | 404.6 |
| 10 | 10 | 1.322 | 188.4 | 0.4927 | 0.15862 | 382.4 |
| 10 | 17 | 1.357 | 189.7 | 0.5000 | 0.15782 | 379.3 |
| 10 | 30 | 1.394 | 191.0 | 0.5079 | 0.15695 | 376.1 |
| 10 | 55 | 1.429 | 191.7 | 0.4875 | 0.15696 | 393.1 |
| 10 | 100 | 1.462 | 192.3 | 0.4674 | 0.15697 | 411.4 |
| 17 | 1 | 5.694 | 189.4 | 0.4333 | 0.03647 | 437.0 |
| 17 | 1.7 | 3.769 | 192.4 | 0.4813 | 0.08284 | 399.8 |
| 17 | 3 | 1.845 | 195.5 | 0.5293 | 0.12921 | 369.3 |
| 17 | 5.5 | 1.612 | 192.0 | 0.5348 | 0.14903 | 359.0 |
| 17 | 10 | 1.379 | 188.5 | 0.5403 | 0.16886 | 349.0 |
| 17 | 17 | 1.357 | 188.2 | 0.5430 | 0.17073 | 346.5 |
| 17 | 30 | 1.336 | 187.8 | 0.5458 | 0.17260 | 344.0 |
| 17 | 55 | 1.241 | 188.0 | 0.5518 | 0.17488 | 340.6 |
| 17 | 100 | 1.147 | 188.1 | 0.5578 | 0.17716 | 337.3 |
| 30 | 1 | 4.456 | 187.3 | 0.5267 | 0.04331 | 355.6 |
| 30 | 1.7 | 3.168 | 189.6 | 0.5624 | 0.08694 | 337.0 |
| 30 | 3 | 1.788 | 192.0 | 0.6007 | 0.13365 | 319.6 |
| 30 | 5.5 | 1.611 | 190.3 | 0.5942 | 0.15653 | 320.3 |
| 30 | 10 | 1.436 | 188.7 | 0.5879 | 0.17909 | 320.9 |
| 30 | 17 | 1.359 | 186.7 | 0.5859 | 0.18352 | 318.6 |
| 30 | 30 | 1.277 | 184.5 | 0.5838 | 0.18825 | 316.1 |
| 30 | 55 | 1.053 | 184.2 | 0.6162 | 0.19283 | 299.0 |
| 30 | 100 | 0.832 | 184.0 | 0.6481 | 0.19735 | 283.8 |
| 55 | 1 | 2.877 | 179.5 | 0.6493 | 0.07236 | 276.4 |
| 55 | 1.7 | 2.121 | 182.7 | 0.6648 | 0.11358 | 274.8 |
| 55 | 3 | 1.365 | 185.9 | 0.6803 | 0.15481 | 273.3 |
| 55 | 5.5 | 1.234 | 184.1 | 0.6693 | 0.17426 | 275.1 |
| 55 | 10 | 1.102 | 182.3 | 0.6582 | 0.19371 | 277.0 |
| 55 | 17 | 1.074 | 182.3 | 0.6532 | 0.19655 | 279.1 |
| 55 | 30 | 1.045 | 182.3 | 0.6483 | 0.19940 | 281.2 |
| 55 | 55 | 0.913 | 181.6 | 0.6574 | 0.20028 | 276.2 |
| 55 | 100 | 0.781 | 180.8 | 0.6665 | 0.20117 | 271.3 |

**Table S1.** (continued)

| $k_{TH}$ | $c_{TH}$ | $k_{med}$ | $U_{med}$ | $g_k$ | $g_U$ | $U_{med}/g_k$ |
|---|---|---|---|---|---|---|
| 100 | 1 | 1.298 | 171.7 | 0.7720 | 0.10141 | 222.4 |
| 100 | 1.7 | 1.126 | 175.6 | 0.7662 | 0.13742 | 229.2 |
| 100 | 3 | 0.943 | 179.9 | 0.7600 | 0.17597 | 236.6 |
| 100 | 5.5 | 0.854 | 177.9 | 0.7442 | 0.19226 | 239.0 |
| 100 | 10 | 0.768 | 175.9 | 0.7285 | 0.20832 | 241.5 |
| 100 | 17 | 0.790 | 177.9 | 0.7209 | 0.20939 | 246.8 |
| 100 | 30 | 0.814 | 180.0 | 0.7128 | 0.21054 | 252.6 |
| 100 | 55 | 0.771 | 178.9 | 0.6988 | 0.20775 | 256.0 |
| 100 | 100 | 0.729 | 177.7 | 0.6849 | 0.20499 | 259.5 |

# Additional Information

**Data Accessibility**
The code and data are in supplementary material.

**Declaration of AI use**
I have not used AI-assisted technologies in creating this article.

**Authors' Contributions**
T.K.: conceptualization, data curation, formal analysis, investigation, methodology, software, validation, visualization, writing—original draft, writing—review & editing. Y.T.: conceptualization, methodology. M.R.H.: conceptualization, methodology.

**Competing Interests**
I have no competing interests.

**Funding**
This work was supported by the JSPS Topic-Setting Program to Advance Cutting-Edge Humanities and Social Sciences Research Grant Number JPJS00122679495.